\documentclass{article}

\usepackage{times}
\usepackage{helvet}
\usepackage{courier}

\usepackage{graphicx,pstricks}
\usepackage{moreverb}
\usepackage{subfigure}
\usepackage{epsfig}
\usepackage{subfigure}
\usepackage{setspace}

\usepackage{pgfplots}
\usepgfplotslibrary{polar}
\usepgflibrary{shapes.geometric}
\usetikzlibrary{calc}

\usepackage{enumerate}
\usepackage{amsfonts}
\usepackage{amssymb}
\usepackage{amsthm}
\usepackage{graphicx}
\usepackage{amsmath}
\usepackage{array}

\newtheorem{theorem}{Theorem}[]
\newtheorem{lemma}[]{Lemma}

\newtheorem{definition}[theorem]{Definition}

\newcommand{\mytie}{\mathrel{\rhd\mspace{-9mu}<}}

\pgfplotsset{my style/.append style={axis x line=bottom, axis y line=left, xlabel={degree bound for $\mathcal{F}$}, ylabel={$c$}}}

\title{Extended
Hardness Results for\\
 Approximate Gr\"{o}bner Basis Computation}
\author{G. Spencer \thanks{Smith College. Northampton, MA 01063. USA. Phone: (607)351-3035 .\textit{Early stages of this work were supported by the Mathematics Research Communities Program of the AMS.}}}
\date{October 2015}

\usepackage{natbib}
\usepackage{graphicx}

\begin{document}

\maketitle

\begin{abstract}
Two models were recently proposed to explore the robust hardness of Gr\"{o}bner basis computation. Given a polynomial system, both models allow an algorithm to selectively ignore some of the polynomials: the algorithm is only responsible for returning a Gr\"{o}bner basis for the ideal generated by the remaining polynomials. For the $q$-Fractional Gr\"{o}bner Basis Problem the algorithm is allowed to ignore a constant $(1-q)$-fraction of the polynomials (subject to one natural structural constraint). Here we prove a new strongest-parameter result: even if the algorithm is allowed to choose a $(3/10-\epsilon)$-fraction of the polynomials to ignore, and need only compute a Gr\"{o}bner basis with respect to some lexicographic order for the remaining polynomials, this cannot be accomplished in polynomial time (unless $P=NP$). This statement holds even if every polynomial has maximum degree 3. Next, we prove the first robust hardness result for polynomial systems of maximum degree 2: for the $q$-Fractional model a $(1/5-\epsilon)$ fraction of the polynomials may be ignored without losing provable NP-Hardness. Both theorems hold even if every polynomial contains at most three distinct variables. Finally, for the Strong $c$-partial Gr\"{o}bner Basis Problem of De Loera et al. we give conditional results that depend on famous (unresolved) conjectures of Khot and Dinur, et al.

\end{abstract}

\noindent \textbf{Keywords:}
\textbf{AMS Subject Classification: 68Q17}\\
\noindent \textbf{Keywords:} Computational algebra, Gr\"{o}bner basis, Complexity, Hardness of approximation, Satisfiability, Theoretical Computer Science

\section{Introduction}
Gr\"{o}bner basis computation is a classic problem in computational algebra (see \cite{sturmfels2005grobner}, \cite{Cox}, or \cite{Laur} for details). A Gr\"{o}bner basis for the polynomial ideal $\langle \mathcal{F}\rangle$ is simply a special set of polynomial generators for $\langle \mathcal{F}\rangle$ with the property that the leading terms of the Gr\"{o}bner basis elements generate all leading terms of the polynomial ideal $\langle \mathcal{F}\rangle$. In contrast, a general set of polynomial generators $\mathcal{F}$ for $\langle \mathcal{F}\rangle$ may not have leading terms with this property: it could be that some leading terms of $\langle \mathcal{F}\rangle$ are obtained through cancellation of leading terms for combinations of polynomials from $\mathcal{F}$.

It is well-known that Gr\"{o}bner basis computation with respect to a lexicographic order is not possible in polynomial-time (unless $P=NP$).  This is obvious because polynomial systems can be used to tidily encode the exact solutions of number of NP-Hard combinatorial optimization problems. Using standard results from elimination theory, once a Gr\"{o}bner basis for a lexicographic order is in hand, an optimal solution can be quickly computed (see \citep{Cox}). The natural polynomial system for the Minimum Vertex Cover Problem (which is NP-Hard) shows that this NP-Hardness for Gr\"{o}bner basis computation holds even when the polynomial system has maximum degree 2.

Despite these general hardness results, many algebraists seem to hope that if the problem of Gr\"{o}bner basis computation is restricted to polynomial systems with a few nice properties, then the problem may be efficiently solvable (in the traditional polynomial-time sense). Indeed, empirical experience running the most famous algorithm for the problem (Buchberger's Algorithm \cite{buch}) makes it tempting to think that addressing polynomials via some clever ordering, or choosing a variable or term ordering based on the input, could somehow resolve various difficulties (intermediate polynomials of very high degree, etc). Unfortunately, recent results of De Loera et al. \cite{deloera} and Rolnick and Spencer \cite{RS} use results from combinatorial optimization and complexity to prove that even \textit{approximate solutions} to Gr\"{o}bner Basis queries cannot be efficiently obtained for lexicographic orders, even for intensely-restricted polynomial-system inputs.\footnote{Notably, their results and ours allow the algorithm to select any lexicographic order that is convenient. Questions about robust hardness of Gr\"{o}bner basis computation with respect to other widely-studied term orders remain completely open.}

In particular, many combinatorial optimization problems are not only \textit{hard to solve exactly in polynomial time}, they are \textit{hard to solve \textbf{even approximately} in polynomial time}. In what meaningful sense might a Gr\"{o}bner basis query be hard to solve \textit{even approximately}? To explore this idea, in \cite{deloera}, De Loera et al. introduced the $c$-Partial Gr\"{o}bner Basis Problem, and later Rolnick and Spencer introduced the $q$-Fractional Gr\"{o}bner Basis Problem \cite{RS}. Both models allow an algorithm to selectively ignore some of the polynomials in the input subject to a structural constraint on the ignored set: the algorithm is only responsible for returning a Gr\"{o}bner Basis for the ideal generated by  the remaining polynomials. The precise definitions of these problems will appear in the following sections. Existing hardness results for the Strong $c$-Partial Gr\"{o}bner Basis Problem have been based on hardness-of-approximation results for graph coloring problems (see \citep{deloera} and \citep{RS}). Existing hardness results for the $q$-Fractional Gr\"{o}bner Basis Problem are based on hardness-of-approximation results for the classic Max-3SAT Problem (see \citep{RS}). We shall be more explicit about these existing results after precisely defining the models.\\


\noindent \textbf{Our contribution.} 
In this paper we give a new best result for the $q$-Fractional Gr\"{o}bner Basis Problem based on the standard assumption that $P\neq NP$. Our reduction uses a 2014 hardenss-of-approximation result of H{\aa}stad for the Max Not-2 Problem.

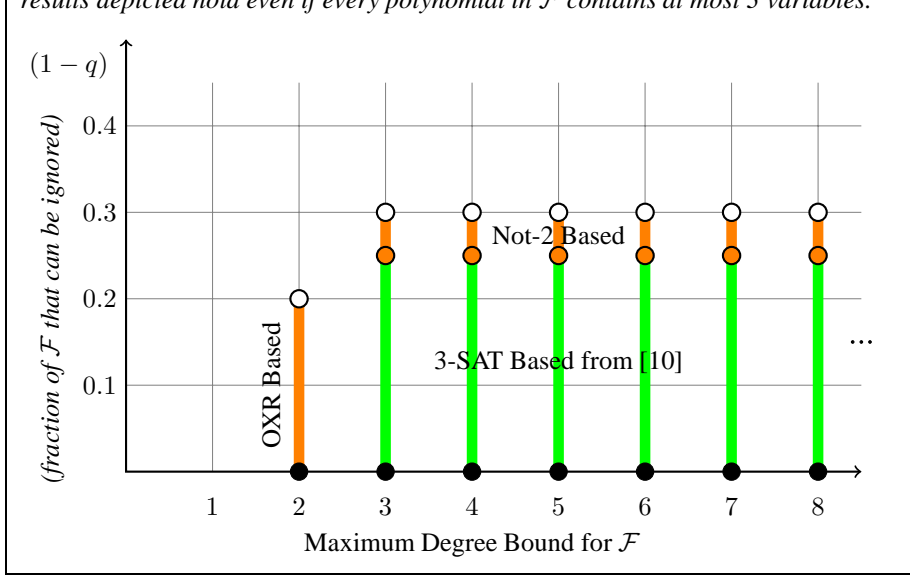
\begin{figure}[h!]\label{graphofq}
\fbox{
\begin{minipage}{11.8 cm}
\caption{\textbf{Values of $(1-q)$ for which the $q$-Fractional Gr\"{o}bner Basis Problem is NP-Hard for lexicographic orders.} \textit{NP-Hardness of the classic Gr\"{o}bner Basis Problem is depicted by the black points on the horizontal axis. The main theorem from Rolnick and Spencer \citep{RS} is depicted in green. Our results overlap \citep{RS}: parameter values covered by our extensions but not by \cite{RS} are depicted in orange. All results depicted hold even if every polynomial in $\mathcal{F}$ contains at most 3 variables.}}

\vspace{3mm}

\begin{tikzpicture}[scale=1.15]
\draw[help lines] (0,0) grid (8.5,4.5);
\draw [ orange, line width=4] (2,0) -- (2,2);

\draw [green, line width=4] (3,0) -- (3,2.5);
\draw [green, line width=4] (4,0) -- (4,2.5);
\draw [green,line width=4] (5,0) -- (5,2.5);
\draw [green, line width=4] (6,0) -- (6,2.5);
\draw [green,line width=4] (7,0) -- (7,2.5);
\draw [green, line width=4] (8,0) -- (8,2.5);

\draw [orange, line width=4] (3,2.5) -- (3,3);
\draw [orange, line width=4] (4,2.5) -- (4,3);
\draw [orange, line width=4] (5,2.5) -- (5,3);
\draw [orange, line width=4] (6,2.5) -- (6,3);
\draw [orange, line width=4] (7,2.5) -- (7,3);
\draw [orange, line width=4] (8,2.5) -- (8,3);

\draw [fill=black] (2,0) circle [radius=0.1];
\draw [fill=black] (3,0) circle [radius=0.1];
\draw [fill=black] (4,0) circle [radius=0.1];
\draw [fill=black] (5,0) circle [radius=0.1];
\draw [fill=black] (6,0) circle [radius=0.1];
\draw [fill=black] (7,0) circle [radius=0.1];
\draw [fill=black] (8,0) circle [radius=0.1];

\draw [fill=black] (8.4,1.5) circle [radius=0.015];
\draw [fill=black] (8.5,1.5) circle [radius=0.015];
\draw [fill=black] (8.6,1.5) circle [radius=0.015];

\draw [fill=orange,thick] (3,2.5) circle [radius=0.1];
\draw [fill=orange,thick] (4,2.5) circle [radius=0.1];
\draw [fill=orange,thick] (5,2.5) circle [radius=0.1];
\draw [fill=orange,thick] (6,2.5) circle [radius=0.1];
\draw [fill=orange,thick] (7,2.5) circle [radius=0.1];
\draw [fill=orange,thick] (8,2.5) circle [radius=0.1];

\draw [fill=white,thick] (3,3) circle [radius=0.1];
\draw [fill=white,thick] (4,3) circle [radius=0.1];
\draw [fill=white,thick] (5,3) circle [radius=0.1];
\draw [fill=white,thick] (6,3) circle [radius=0.1];
\draw [fill=white,thick] (7,3) circle [radius=0.1];
\draw [fill=white,thick] (8,3) circle [radius=0.1];

\draw [fill=white,thick] (2,2) circle [radius=0.1];

\draw [<->, thick] (0,5) -- (0,0) -- (8.5,0);

\node[align=center, below] at (1,-.2) {$1$};
\node[align=center, below] at (2,-.2) {$2$};
\node[align=center, below] at (3,-.2) {$3$};
\node[align=center, below] at (4,-.2) {$4$};
\node[align=center, below] at (5,-.2) {$5$};
\node[align=center, below] at (6,-.2) {$6$};
\node[align=center, below] at (7,-.2) {$7$};
\node[align=center, below] at (8,-.2) {$8$};

\node[align=center, below] at (4,-.6)%
{Maximum Degree Bound for $\mathcal{F}$};

\node[label=left:\rotatebox{90}{\textit{(fraction of $\mathcal{F}$ that can be ignored)}}] at (-0.5,2) {};

\node[align=center, left] at (-0.1,4.7) {$(1-q)$};

\node[align=center, left] at (0,1) {$0.1$};
\node[align=center, left] at (0,2) {$0.2$};
\node[align=center, left] at (0,3) {$0.3$};
\node[align=center, left] at (0,4) {$0.4$};


\node[align=center, below] at (5,1.5)%
{3-SAT Based from \cite{RS}};

\node[align=center, below] at (5,2.95)%
{Not-2 Based};

\node[label=left:\rotatebox{90}{OXR Based}] at (2,1) {};

\end{tikzpicture}
\end{minipage}
}
\end{figure}

At a high level, our reduction is very similar to the original Max-3SAT-based result  in \citep{RS}: Approximate Gr\"{o}bner Basis computation is used to construct a partial assignment that is perfect on some fraction of the input, and then a random assignment procedure with modest expected quality can be used to supplement the fraction of satisfied clauses beyond what is allowed by known hardness-of-approximation bounds.
To beat the result based on Max 3SAT, we turn to a different form of logical predicate where a much stronger hardness-of-approximation bound holds. Unfortunately, switching from the disjunctions of 3-SAT to Not-2 predicates sacrifices a key property of the varieties of polynomial systems constructed in \citep{RS}. Significant additional work is required to beat \citep{RS}: the constructed polynomial system is more complex, and the quality delivered by both the Gr\"{o}bner Basis-based and coin-flip-based portions of the assignment rely on structural properties revealed by careful pre-processing of satisfiable Max Not-2 instances. We prove that even for polynomial systems of maximum degree 3 in which each polynomial contains at most 3 variables, the $(7/10+\epsilon)$-Fractional Gr\"{o}bner Basis Problem with respect to lexicographic orders is NP-Hard for any $\epsilon>0$.

Resolving the issues associated with Not-2 predicates, we realized that a highly similar proof could be based on an older 2001 hardness result of H{\aa}stad for the Max-OXR predicate problem \citep{hast01}. Our OXR-based result gives a weaker parameter ($q=4/5+\epsilon$) for the $q$-Fractional model, but the result holds even 
for polynomial systems with maximum total degree 2. This is the first robust hardness result for degree 2 (matching the degree bound for ordinary NP-Hardness of the traditional Gr\"{o}bner Basis computation). Anecdotally, we have found that algebraists seem quite surprised that a robust-hardness notion holds for polynomial systems of maximum degree 2. Further, it seems unlikely that a ``max-degree 2" result can be shown for the $c$-partial model of Robust Hardness of Gr\"{o}bner Basis Computation.
Our new contributions for the $q$-Fractional problem are summarized in Figure \ref{graphofq}.

Finally, we point out 2 new conditional results for the Strong $c$-partial Gr\"{o}bner Basis Problem of De Loera et al.  In \citep{RS}, Rolnick and Spencer's reduction for the $c$-partial model applied an older (non-conditional) NP-Hardness result of Lund and Yanakakis \citep{Lundyan} for the ApproxColoring$(q,Q)$ Problem.
In 2009, Dinur, Mossel and Regev \citep{dinur} showed that Khot's $2\leftrightarrow 2$ conjecture implies surprising strong conditional results for the ApproxColoring$(q,Q)$ Problem. Further, Dinur, Mossel and Regev propose a new label-cover-related conjecture that implies an even stronger conditional result for ApproxColoring$(q,Q)$. In the reduction from \citep{RS}, we substitute Dinur et al.s conditional results directly in the place of the NP-Hardness result of Lund and Yanakakis: the paramaters that result for the Strong $c$-partial problem seem very surprising. As with both conditional results of Dinur et al., our conditional results either suggest deep robust hardness of a classic problem, or, by extending Khot's conjectures into areas where there are many tools (graph coloring and computational algebra respectively), perhaps supply useful directions for possible proofs by contradiction.

\section{Extended Results for $q$-Fractional Gr\"{o}bner Basis Computation}
Rolnick and Spencer showed that the  $(\frac{3}{4}+\epsilon)$-Fractional Gr\"{o}bner Basis Problem is NP-Hard for lexicographic orders. 
That result was proved by reducing from the \textit{Max-3SAT Problem for satisfiable instances}. An input to the \textit{3SAT} problem consists of a set of logical clauses over a set of literals, where each clause is a disjunction of at most 3 literals (or negations of literals). 
The objective in \textit{Max-SAT} problems is to choose a truth assignment for the literals that satisfies the largest fraction of the set of clauses. When we study \textit{Max-SAT for satisfiable instances} we are guaranteed that a truth assignment which satisfies all clauses exists: H{\aa}stad's celebrated result for \textit{Max-3SAT} shows that even with such a guarantee of satisfiability it is NP-Hard to produce a truth assignment satisfying a $(7/8+\epsilon)$ fraction of the clauses (for any $\epsilon>0$).  Rolnick and Spencer define an polynomial system based on an arbitrary instance of \textit{Max-3SAT for satisfiable instances} and show how the ability to efficiently solve the $q$-Fractional Gr\"{o}bner Basis Problem well for a lexicographic order can be used to construct a solution of quality that violates H{\aa}stad's bound. 

First, in Section \ref{not2sec}, we give a
new reduction for the $q$-Fractional Gr\"{o}bner Basis Problem  that is similar in character to that in \citep{RS}, but rather than focusing on Satisfiability for clauses which are disjunctions
(perhaps one of the best known problems in complexity theory),
we turn to a different form of predicate over binary $\{0,1\}$ variables. We reduce from \textit{Max Not-2} (``Not two") \textit{for satisfiable instances} for predicates of arity 3. ``Not two" predicates accept any sum that is not 2, and reject otherwise. We apply a recent $(5/8+\epsilon)$ hardness-of-approximation result for satisfiable instances for this problem also due to H{\aa}stad (which holds
provided that $P\neq NP$).\footnote{Before we became aware H{\aa}stad's 2014 paper, we had proved a conditional version of our main contribution here based on an '09 conditional result of O'Donnell and Wu.  O'Donnell and Wu's version gives the same inapproxamability factor as \citep{hast14} but relies on the assumption that Khot's $d-to-1$ Conjecture holds for some finite $d$.}
Unlike in \citep{RS}, for \textit{Max-Not-2} the natural set of ``predicate polynomials" unfortunately yields a variety containing many points that can't be interpreted as solutions to our combinatorial problem. To get a better correspondence, we are forced to include an additional family of polynomials in our constructed polynomial system, however, by tailoring the \textit{Max-Not-2} instance in an initial stage, we can bound the relative size of this family. As in \citep{RS} we analyze a random assignment procedure on the ignored literals: the Not-2 predicates require careful case analysis to achieve a high enough rate of expected satisfaction, and again we rely on a property made possible by our initial stage of Not-2-instance tailoring. 

Next, in Section \ref{deg2sec}, we reuse the high-level structure of our Max-Not-2 proof, but now leveraging the Max-OXR Problem (yet another form of logical predicate). Here we use an earlier 2001 hardness result of H{\aa}stad that appears weaker in its inapproxamability parameter. The form of the OXR predicates is worth the sacrifice: these predicates naturally yield polynomial systems of maximum total degree 2. Further, these polynomial systems are still sparse in the sense that every polynomial contains at most 3 variables. Thus, we give the first robust hardness result for Gr\"{o}bner Basis computation in polynomial systems of max degree 2: the $(4/5+\epsilon)$-Fractional Gr\"{o}bner Basis Problem is NP-Hard. 

\subsection{Extended Hardness Result: A $(\frac{3}{10}-\epsilon)$ Fraction of $\mathcal{F}$ May be Selectively Ignored} \label{not2sec}

Recall the definition from \citep{RS}. For consistency, we duplicate their notation exactly. For polynomial system $\mathcal{F}$, and subset of variables $Y$, let $\mathcal{F}_Y$ denote the subset of polynomials from $\mathcal{F}$ which contain at least one variable from $Y$.

\begin{definition}
\textbf{$q$-Fractional Gr\"{o}bner Basis Problem}. Given as input a set of polynomials $\mathcal{F}$ over a set of variables $X$, for specified $q\in [0,1]$, output the following:
\begin{itemize}
\item $X'\subseteq X$, such that $|\mathcal{F}_{X'}|\leq (1-q)|\mathcal{F}|$.
\item A Gr\"{o}bner Basis for $\mathcal{F}\backslash \mathcal{F}_{X'}$.
\end{itemize}
\end{definition}

\noindent For $q=1$, this is exactly the traditional Gr\"{o}bner Basis problem for $\mathcal{F}$. As described in \citep {RS}, the set $X'$ corresponds to a set of variables \textit{chosen by the algorithm to be ignored}: all the polynomials containing variables from $X'$ are ignored, and the algorithm need only compute a Gr\"{o}bner Basis for the remaining set of polynomials $\mathcal{F}\backslash \mathcal{F}_{X'}$.

\begin{theorem} \textbf{Extending Robust Hardness.} \label{extfrachard}
Assume that we are working over a polynomial ring $\mathbb{K}[x_1,x_2,x_3...x_n]$. For any $\epsilon>0$: there is no polynomial-time algorithm $\mathcal{A}$ that solves the $(7/10+\epsilon)$-Fractional Gr\"{o}bner Basis Problem with respect to any lexicographic order (unless P = NP). This statement holds even when $\mathcal{F}$ has maximum degree 3, and each polynomial from $\mathcal{F}$ contains at most 3 variables.
\end{theorem}

We prove Theorem \ref{extfrachard} by reducing from the \textit{Max-Not-2 Problem for satisfiable instances of arity 3}. The input to the Max-Not-2 Problem is a set of logical predicates $\mathcal{P}$ over a set of literals $\mathcal{L}$. Specifying \textit{arity 3} means that each predicate contains at most 3 signed literals (a ``signed literal" is just a literal in either negated or positive form), e.g. $(l_i,l_j,\neg l_k)$ where $l_i, l_j,l_k\in \mathcal{L}$. For a truth assignment to the literals, a predicate is ``satisfied" if the number of its signed literals that are true is not 2. If exactly 2 of its signed literals are true, then the predicate is not satisfied. For example, the predicate $(l_i,l_j,\neg l_k)$ is satisfied by a truth assignment where $l_i$ is true, $l_j$ is true, and $l_k $ is false (all three of the signed literals in the predicate are true for this truth assignment). On the other hand, consider a truth assignment in which $l_i$ is true, $l_j$ is true, and $l_k$ is true: for this truth assignment exactly 2 of the signed literals in the predicate are true, so the predicate is not satisfied. The objective for the Max-Not-2 Problem is to compute a truth assignment that satisfies the highest possible fraction of predicates in $\mathcal{P}$. To say that we consider the problem on \textit{satisfiable instances} means that we receive the input together with a guarantee that there exists some truth assignment for $\mathcal{L}$ which satisfies every predicate in $\mathcal{P}$.

Now consider the recent result of H{\aa}stad:

\begin{theorem} \label{hastad}(H{\aa}stad, '14)
For any $\delta>0$, given a satisfiable instance of Max Not-2 of arity 3, there is no polynomial-time algorithm to find a truth assignment that satisfies a $(\frac{5}{8}+\delta)$-fraction of the predicates (unless $P=NP$). 
\end{theorem}

As is standard (and will be useful in describing and analyzing our reduction), we can equivalently take an algebraic view of Max Not-2 and consider each literal $l_i$ to be a $\{0,1\}$ variable $x_i$ (where $x_i=1$ corresponds to $l_i$ true, and $x_i=0$ corresponds to $l_i$ false), and translate each predicate into a sum.  If a predicate contains a signed literal in positive form, a positive copy of the corresponding variable is added. If a predicate contains a signed literal in negated form, a term is added in which the corresponding variable is subtracted from 1. For example, the predicate $(l_i,l_j,\neg l_k)$ becomes the sum
\[
x_i+x_j+(1-x_k).
\]
It is easy to check that the original predicate is satisfied exactly when its corresponding sum is not 2 (and hence has total value 0, 1, or 3). The language of our proof will deal with an input specified in terms of such sums for predicates and variables for literals. Describing an input for the Max-Not-2 Problem of arity 3 we will say each predicate \textit{has at most three acceptable totals}, and exactly \textit{one unacceptable total}.\footnote{If the predicate has fewer than three signed literals, the number of achievable acceptable totals maybe be less than 3.}

Finally, before starting the main proof, we mention that as in \citep{RS} our reduction will use the crucial fact that if we possess a Gr\"{o}bner Basis for a polynomial system with respect to a lexicographic order, then a point in the variety can be computed efficiently by iteratively eliminating the variables one at a time. In fact, our situation will be simpler than in \citep{RS} because the variety of the polynomial system defined in our reduction is finite so that all partial solutions extend during the elimination procedure. These classic results in elimination theory are covered in the textbook of Cox, Little and O'Shea \cite{Cox}.\\

\noindent \textit{Proof (Theorem \ref{extfrachard}):} Suppose, for the sake of contradiction, that the $\mathcal{A}$ asserted in Theorem \ref{extfrachard} does exist with $q=(7/10+\epsilon)$ for some fixed $\epsilon>0$. Given an arbitrary satisfiable input $(\mathcal{P}, \mathcal{L})$ of the \textit{Max-Not-2 Problem of arity 3} we compute an assignment of forbidden quality in polynomial time as follows. Our assignment will be determined over the course of three stages. \\

\noindent \textbf{Stage 1. Instance Tailoring.} We remove some predicates and literals from $(\mathcal{P}, \mathcal{L})$ so that certain useful properties hold. 

Iterate through the predicates in $\mathcal{P}$ one at a time.  If $p\in \mathcal{P}$ has strictly more than one signed literal corresponding to a single literal $l_i$, then update $(\mathcal{P}, \mathcal{L})$ according to which of the following cases applies:
\begin{enumerate}
\item If $p$ contains three identical signed literals, then $p$ is trivially satisfied (every assignment for $\mathcal{L}$ satisfies $p$). Remove $p$ from $\mathcal{P}$.
\item Otherwise, if $p$ contains exactly 2 identical signed literals indexed by $i$ then:
\begin{enumerate}
\item If $p$'s third signed literal is the other form of $l_i$: every satisfying assignment for $(\mathcal{P}, \mathcal{L})$ must cause $p$ to contain exactly 1 true literal (the only alternative is 2, which can't be). Thus, we know unequivocally the value $x_i$ must take in every satisfying assignment. Substitute the forced value of $x_i$ into every predicate containing a signed form of literal $l_i$. Say $x_i$ has been permanently fixed. Remove the literal indexed by $i$ from $\mathcal{L}$, and remove $p$ from $\mathcal{P}$. 
\item If $p$ has no third signed literal: every satisfying assignment for $(\mathcal{P}, \mathcal{L})$ must cause $p$ to contain exactly 0 true literals (the only alternative is 2, which can't be). Thus, we know unequivocally the value $x_i$ must take in every satisfying assignment. Substitute the forced value of $x_i$ into every predicate containing a signed form of literal $l_i$. Say $x_i$ has been permanently fixed. Remove the literal indexed by $i$ from $\mathcal{L}$, and remove $p$ from $\mathcal{P}$. 

\item If $p$'s third signed literal corresponds to some other index $j$ where $x_j$ has not yet been fixed, do nothing.

\item If $p$'s third signed literal originally corresponded to some other index $j$ where $x_j$ has already been permanently fixed to 0 or 1, then do nothing. 

\end{enumerate}
\item Otherwise, it must be that $p$ contains 2 signed literals corresponding to the same index $i$, but with opposing forms (one negated and one positive).  Then:
\begin{enumerate}

\item If $p$'s third signed literal is also for index $i$, then it must be identical to one of $p$'s opposing-signed literals: this was already covered in sub-case 2(a) above.
\item  If $p$'s third signed literal corresponds to some other index $j$ for which $x_j$ has not yet been fixed: since $p$'s opposing literals have sum 1, we know unequivocally the value $x_j$ must take in every satisfying assignment. Substitute the forced value of $x_j$ into every predicate containing a signed form of literal $l_j$. Say $x_j$ has been permanently fixed. 
 Remove the literal indexed by $j$ from $\mathcal{L}$, and remove $p$ from $\mathcal{P}$. 
\item If $p$'s original third signed literal corresponded to some other index $j$ for which $x_j$ has already been permanently fixed to 0 or 1: since variables are only fixed to values we know they must take in every satisfying assignment, $x_j$ must have been fixed so that the term in which it appeared in predicate $p$ had value 0.\footnote{If the term in which $x_j$ appeared had value 1, then since the sum of the terms for the opposing form signed literals involving $l_i$ is always 1, no choice of $x_i$ would satisfy $p$, and this would contradict the satisfiability of $(\mathcal{P}, \mathcal{L})$. } Thus, $p$ will be satisfied by any assignment for $x_i$. Remove $p$ from $\mathcal{P}$.

\item If $p$ never had a third signed literal:  $p$ is trivially satisfied (every assignment for $\mathcal{L}$ satisfies $p$). Remove $p$ from $\mathcal{P}$.

\end{enumerate}
\end{enumerate}

Call the set of all literals fixed during this procedure $\mathcal{L}^f$, and the set of all predicates removed from the original $\mathcal{P}$ by $\mathcal{P}^r$.
After executing the above procedure for every $p\in \mathcal{P}$, observe that $(\mathcal{P}, \mathcal{L})$ has the following property.  \\

\noindent \textbf{Property 1.} \textit{Any predicate $p'\in \mathcal{P}$ that has multiple occurrences of the same literal must have a very specific form: either $p'$ contains two identical signed literals and a third signed literal corresponding to a different index, or  $p'$ contains two identical signed literals and a third term whose value has been permanently fixed to either 0 or 1. These forms correspond to sub-cases 2(c) and 2(d) above: in all other sub-cases, $p$ was removed from $\mathcal{P}$.}\\

Further observe that variables were only permanently fixed (and removed from the set of literals) when we could reason unequivocally about the value they must take in every satisfying assignment. Thus, since the original $(\mathcal{P}, \mathcal{L})$ was satisfiable, the updated $(\mathcal{P}, \mathcal{L})$ is still satisfiable. Before proceeding, notice also that a predicate was only removed from $\mathcal{P}$ when we could be certain that it would be satisfied by any assignment that extends the partial assignment already constructed for permanently-fixed variables.

Next, we make a final update to  $(\mathcal{P}, \mathcal{L})$. Call any literal $\l\in \mathcal{L}$ which appears in at most one predicate from $\mathcal{P}$ a \textit{Loner Literal}. Call the set of predicates from $\mathcal{P}$ which contain at least one Loner Literal by $\mathcal{P}^l$. We consider temporarily ignoring predicates in $\mathcal{P}^l$ until all non-loner literals have been fixed. Consider an arbitrary partial assignment for $(x_1,...,x_{|\mathcal{L}|})$ that fixes every literal variable except for the \textit{Loner Literals}. If a predicate  $p\in \mathcal{P}$ contains one or more \textit{Loner Literals} (in either positive or negated form), then this arbitrary partial assignment may be easily extended to satisfy $p$: by manipulating the $\{0,1\}$ value of a contained \textit{Loner Literal} at least 2 distinct total sums may be reached for predicate $p$ (this follows from Property 1). At most one of these total sums can be equal to 2, and the other must be some \textit{acceptable total} for $p$ (such that $p$ is satisfied). Since the \textit{Loner Literals} in $p$ appear in no other predicates (by definition), their value may be fixed one-by-one in this way to satisfy all predicates in $\mathcal{P}^l$. Since we can successfully satisfy them at the end, we ignore such predicates for now: if a predicate contains a \textit{Loner Literal}, remove that predicate from $\mathcal{P}$. Next remove all \textit{Loner Literals} from $\mathcal{L}$. 

Notice that these removing the predicates in $\mathcal{P}^l$ may have caused some additional literals to become \textit{Loner Literals}. Successively remove additional rounds of \textit{Loner}-containing predicates and \textit{Loner Literals}. Mark each \textit{Loner Literal} by the round in which it was removed: once we have created a partial assignment for the remaining system we will fix the values of the \textit{Loner literals} in an order that reverses the order in which they were removed from $\mathcal{L}$. Such an ordering ensures that as \textit{Loners} from each round are returned to the instance, the argument made in the previous paragraph about how to choose their value will always apply.

We now have the $(\mathcal{P}, \mathcal{L})$ that we will argue about for the remainder of the reduction.  We make a few observations before starting Stage 2. Property 1 still holds, and as a result of the \textit{Loner}-removal and \textit{Loner}-containing-predicate removal process, we have:\\

\noindent \textbf{Property 2.} \textit{After the updates in Stage 1, Every literal $l\in \mathcal{L}$ appears in some form (negated or positive) in at least two predicates from $\mathcal{P}$.}\\

\noindent Property 2 immediately implies a fact crucial to our subsequent analysis:

\begin{lemma} \label{twofifths} An instance of \textit{Max Not-2} of arity $3$ over literals $\mathcal{L}$ and predicates $\mathcal{P}$ for which each literal appears in at least 2 predicates has $|\mathcal{P}|\geq \frac{2}{5}(|\mathcal{P}|+|\mathcal{L}|)$.
\end{lemma}

\noindent\textit{Proof of Lemma:} If there are $|\mathcal{L}|$ literals, and each appears at least twice, then there must be at least $2|\mathcal{L}|$ appearances of literals. Each predicate contains at most 3 appearances of literals, so at minimum there are $2|\mathcal{L}|/3$ predicates in $\mathcal{P}$:
\begin{align*}
|\mathcal{P}|&\geq \frac{2}{3} |\mathcal{L}|\\
\frac{|\mathcal{P}|}{|\mathcal{P}|+|\mathcal{L}|}&\geq \frac{2}{3} \Big(\frac{|\mathcal{L}|}{|\mathcal{P}|+|\mathcal{L}|}\Big)\\
\frac{|\mathcal{P}|}{|\mathcal{P}|+|\mathcal{L}|}&\geq \frac{2}{3} \Big(1-\frac{|\mathcal{P}|}{|\mathcal{P}|+|\mathcal{L}|}\Big)\\
\frac{5}{3}\Big(\frac{|\mathcal{P}|}{|\mathcal{P}|+|\mathcal{L}|}\Big)&\geq \frac{2}{3}\\
|\mathcal{P}|&\geq \frac{2}{5}(|\mathcal{P}|+|\mathcal{L}|).
\end{align*}
\begin{flushright}$\Box.$\\
\end{flushright}

Consider the two types of predicates removed from the original instance during Stage 1.  Each predicate in $\mathcal{P}^r$ (those removed in the first part of Stage 1) is already guaranteed to be satisfied by any extension of the partial assignment that has been  permanently fixed on $\mathcal{L}^f$. Further, for any partial assignment for literals now remaining in $\mathcal{L}$ that leaves the values of \textit{Loner Literals} (from every round) unassigned, 
each predicate removed for containing a \textit{Loner Literal} (in any round) can be efficiently satisfied by appropriate choices for the \textit{Loner Literals}. 

Since $100\%$ of the predicates removed from $\mathcal{P}$ in Stage 1 can be satisfied in one of these two ways (in polynomial time), any fraction of the predicates we can satisfy for the remaining instance $(\mathcal{P}, \mathcal{L})$ will be a lower bound on the fraction of the original predicates that are satisfied by our assignment method. Thus, to get a contradiction, it is sufficient to show that for our remaining instance we can satisfy a $(\frac{5}{8}+\epsilon)$ fraction of the remaining predicates $\mathcal{P}$. We construct such an assignment for the remaining literals in $\mathcal{L}$ over two stages.\\

\noindent \textbf{Stage 2. Polynomial Encoding.} Recall the algebraic view of $(\mathcal{P}, \mathcal{L})$. For ease of exposition, rename the $x_i$ so that those that remain in $\mathcal{L}$ are indexed consecutively from 1 to $|\mathcal{L}|$.

Define a polynomial system based on $(\mathcal{P}, \mathcal{L})$ as follows. Create a variable $y_i$ corresponding to the $i$th literal of $\mathcal{L}$. Denote this new set of variables by $Y$. These $y_i$ will replicate the $x_i$ in the algebraic view of Max Not-2: for each $i\in \{1,2,...,|\mathcal{L}|\}$ create a polynomial $y_i(1-y_i)$. This gives a set of $|\mathcal{L}|$ ``literal polynomials" whose mutual roots are exactly $\{0,1\}^{|\mathcal{L}|}$. 

Next, create a polynomial corresponding to each predicate. In the algebraic view of Max Not-2, each predicate $p$ corresponds to a sum with \textit{at most 3 acceptable total outcomes} (some subset of $\{0,1,3\}$ gives the acceptable totals for which the sum corresponds to a satisfied predicate). These sum expressions have been altered in Stage 1 as some of the variable values have been permanently fixed. Each acceptable total for the sum corresponding predicate $p$ will be used to define a linear term, and the product of these linear terms will give the polynomial corresponding to predicate $p$. Each linear term is $p$'s sum (with the fixed variables from Stage 1 substituted in) minus an acceptable total for the sum. For example, recall that the algebraic view of the Not-2 predicate $(l_i,l_j,\neg l_k)$ is the sum
\[
x_i+x_j+(1-x_k)\neq 2
\]
If none of $x_i, x_j, x_k$ were fixed in Stage 1, then this yields the following polynomial:
\[
\Big(y_i+y_j+(1-y_k)-0\Big)\Big(y_i+y_j+(1-y_k)-1\Big)\Big(y_i+y_j+(1-y_k)-3\Big)
\]
Since there are at most 3 acceptable totals for $p$'s sum, the polynomial constructed is the product of at most 3 linear terms, and hence has degree at most 3. Since each predicate has at most 3 signed literals, each polynomial will contain at most 3 variables. Notice that when restricted to the Variety for the set of literal polynomials defined above, 0s for $p$'s polynomial correspond exactly to the cases in which $p$'s sum takes on an \textit{acceptable total}. 

All of these observations hold if some of the variables in $p$'s sum were already fixed in Stage 1. For example, if in Stage 1, $x_i, x_j$ were not fixed but $x_k$ was fixed to 0, then $p$'s sum would be $x_i+x_j+(1-0)\neq 2$ and consequently $p$'s polynomial construction would be: 
\begin{align*}
\Big(y_i+y_j+(1-0)-0\Big)\Big(y_i+y_j+(1-0)-1\Big)\Big(y_i+y_j+(1-0)-3\Big)&=\\
\Big(y_i+y_j+1\Big)\Big(y_i+y_j\Big)\Big(y_i+y_j-2\Big).
\end{align*}

This constructs a set of ``predicate polynomials" of size $|\mathcal{P}|$. Let $\mathcal{F}$ denote the system of polynomials containing both the literal polynomials and the predicate polynomials. Notice that every satisfying assignment for $(\mathcal{P},\mathcal{L})$ can be interpreted as a point in the variety defined by $\mathcal{F}$. In particular, since $(\mathcal{P},\mathcal{L})$ is satisfiable, $V(\langle \mathcal F \rangle)$ is non-empty.

Apply algorithm $\mathcal{A}$ to solve the $q$-Fractional Gr\"{o}bner Basis Problem for $\mathcal{F}$ for $q=(7/10+\epsilon)$ for some fixed $\epsilon>0$. Let $Y'$ denote the variables that $\mathcal{A}$ selects to ignore. The set $Y'$ was chosen so that
\begin{align*}
|\mathcal{F}_{Y'}|\leq (1-q)|\mathcal{F}|&\leq (3/10-\epsilon)|\mathcal{F}|\\
&\leq (3/10-\epsilon)(|\mathcal{P}|+|\mathcal{L}|)\\
&\leq (3/10-\epsilon)\Big(\frac{5}{2}|\mathcal{P}|\Big)\\
&\leq \Big(\frac{3}{4}-\frac{5}{2}\epsilon\Big)|\mathcal{P}|
\end{align*}
The third line follows from the second line due to Lemma \ref{twofifths}. 

Let $\mathcal{P}_D$ denote the set of predicate polynomials that are in $\mathcal{F}_{Y'}$, and $\mathcal{P}_R$ denote the set of predicate polynomials that are in $\mathcal{F}\backslash\mathcal{F}_{Y'}$. Clearly $|\mathcal{P}_D|\leq |\mathcal{F}_{Y'}|\leq \Big(\frac{3}{4}-\frac{5}{2}\epsilon\Big)|\mathcal{P}|$, so:
\begin{align}\label{quarter}
|\mathcal{P}_R|= |\mathcal{P}|- |\mathcal{P}_D| \geq |\mathcal{P}|- \Big(\frac{3}{4}-\frac{5}{2}\epsilon\Big)|\mathcal{P}|= \Big(\frac{1}{4}+\frac{5}{2}\epsilon\Big)|\mathcal{P}|
\end{align}

That is, $\mathcal{A}$ computes a Gr\"{o}bner Basis with respect to a lexicographic order for the ideal generated by the polynomials in $\mathcal{F}\backslash \mathcal{F}_{Y'}$, and inequality (\ref{quarter}) says that at least a $(\frac{1}{4}+\frac{5}{2}\epsilon)$ fraction of the predicate polynomials must be in $\mathcal{F}\backslash \mathcal{F}_{Y'}$. The satisfiability of $(\mathcal{P}, \mathcal{L})$ ensured that $V(\langle \mathcal{F}\rangle)$ was non-empty, so $V(\langle \mathcal{F}\backslash \mathcal{F}_{Y'}\rangle)$ is certainly non-empty.
Given the Gr\"{o}bner Basis for $\langle\mathcal{F}\backslash \mathcal{F}_{Y'}\rangle$ with respect to a lexicographic order, a point in the variety of $\langle\mathcal{F}\backslash \mathcal{F}_{Y'}\rangle$ can be efficiently computed via successive elimination of the variables: since the variety is finite (it is a subset of $\{0,1\}^{|Y\backslash Y'|})$, all partial solutions extend, and for each successive variable elimination only 2 options must be checked to find some $y_i$ that works. This point in the variety is a vector $y$ of length $|Y\backslash Y'|$ which is a mutual zero of all polynomials in $\mathcal{F}\backslash \mathcal{F}_{Y'}$. Each entry in the vector corresponds to some literal variable in our Max Not-2 instance: if $y_i$ is 1 in this vector, assign the corresponding literal $x_i$ to be 1, if $y_i$ is 0 in this vector, assign the corresponding literal $x_i$ to be 0.\footnote{Unlike in \citep{RS}, because of the inclusion of the literal polynomials, and the fact that by definition $\mathcal{F}\backslash \mathcal{F}_{Y'}$ contains all the literal polynomials corresponding to variables in $Y\backslash Y'$, this routine does make an assignment for every literal $x_i$ corresponding to a  $y_i \in (Y\backslash Y'$).} 

The vector $y$ is a mutual zero of polynomials in $\mathcal{F}\backslash \mathcal{F}_{Y'}$: substituting $y$ into any predicate polynomial
in $\mathcal{F}\backslash \mathcal{F}_{Y'}$ gives 0. By our construction of the predicate polynomials, this implies that our partial assignment based on $y$ gives an $x$ that yields an \textit{acceptable total} for every predicate polynomial in $\mathcal{F}\backslash \mathcal{F}_{Y'}$. That is, every predicate whose polynomial is in $\mathcal{P}_R$ is satisfied by our partial assignment for $x$.\\

\noindent \textbf{Stage 3. Supplemental Random Assignment.} The literals corresponding to variables in $Y'$ have not yet been assigned values. For these variables, as in \citep{RS}, consider a random independent-fair-coin-flip procedure that assigns $x_i=1$ with probability $\frac{1}{2}$, and $x_i=0$ with probability $\frac{1}{2}$. We will argue that regardless of the partial assignment constructed for $x$ in Stage 2 (and effectively in Stage 1, through the substitution of fixed variable values into the predicates), in expectation this random procedure satisfies at least half of the predicates corresponding to the polynomials in $\mathcal{P}_D$. As in \citep{RS}, such a procedure can be derandomized via the well-known method of conditional expectations to obtain a deterministic assignment algorithm with quality that matches the expected value.

Let $p$ denote a predicate corresponding to a polynomial in $\mathcal{P}_D\subseteq \mathcal{F}_{Y'}$. From the form of $\mathcal{F}_{Y'}$, $p$ contains at least one variable corresponding to a $y_i\in Y'$. Thus, at least one of $p$s signed literals will have truth value determined by the coin-flipping procedure. We analyze the probability that $p$ has an \textit{acceptable total} (and is thus satisfied) at the end of the random assignment procedure. 

The key point in the following case analysis is that $p$ has some \textit{current achieved sum} at the end of Stage 2 (before random coin-flipping-based assignment begins), and for $p$ to be satisfied $p$'s total sum must avoid exactly one \textit{unacceptable total} (namely 2). We'll call the difference of these two values $p$'s \textit{forbidden margin}. For example, if $p$ enters Stage 3 with no signed literals fixed, $p$'s \textit{forbidden margin} will be 2. If $p$ enters Stage 3 with exactly 2 signed literals fixed to be true, and one not yet fixed, then $p$'s \textit{forbidden margin} will be 0.  

First, suppose that all signed literals in $p$ correspond to unique literals. The coin flips for variable values are independent, so the possible probability spaces are as follows.\\

\renewcommand\arraystretch{1.3} 

\begin{tabular}{|l|l|l|} \hline
Number of Un-fixed Signed  & Margin: Additional Signed  & Probability  \\
Literals Entering Stage 3 & Literals True After Stage 3& Distribution  \\
\hline
3 & $\{0,1,2,3\}$& $(\frac{1}{8},\frac{3}{8},\frac{3}{8},\frac{1}{8})$\\ 
\hline
2 & $\{0,1,2\}$& $(\frac{1}{4},\frac{1}{2},\frac{1}{4})$\\ 
\hline
1 & $\{0,1\}$& $(\frac{1}{2},\frac{1}{2})$\\ 
\hline
\end{tabular}

\vspace{4mm}

Notice: regardless of the number of un-fixed signed literals in $p$ entering Stage 3, there is no single outcome whose probability is strictly more than $\frac{1}{2}$.  Thus, regardless of the value of $p$'s \textit{forbidden margin}, the probability that margin is realized is less than or equal to $\frac{1}{2}$. Thus, the probability that $p$ is satisfied (equivalently, that $p$'s total sum is some \textit{acceptable total}) is greater than or equal to $\frac{1}{2}$. 

Now suppose that the signed literals in $p$ do not correspond to unique literals. From \textbf{Property 1} in Stage 1, we have that there are only two possible forms for such a predicate. Again, we argue that in any case, the probability that $p$ reaches an \textit{acceptable total} is at least $\frac{1}{2}$:
\begin{enumerate}
\item Suppose that $p$ is two identical signed literals with a third signed literal that has already been fixed to 0 or 1 in Stage 1. Since $p\in\mathcal{P}_D$, the variable corresponding to the two identical signed literals must be un-fixed entering Stage 3. If the fixed third signed literal has value 0, then the probability that $p$ avoids a total sum of 2 is $\frac{1}{2}$. If the fixed third signed literal has value 1, then the probability that $p$ avoids a total sum of 2 is $1$.

\item Suppose that $p$ is two identical signed literals with a third term that is a signed literal corresponding to an unrelated index. Again $p$ has some well defined \textit{forbidden margin}, and we compile a table of possible probability spaces.

\begin{tabular}{|l|l|l|} \hline
State Entering Stage 3 & Margin: Additional Signed  & Probability  \\
 & Literals True After Stage 3& Distribution  \\
\hline
(pair fixed at 0 or 2, single un-fixed)& $\{0,1\}$&$(\frac{1}{2},\frac{1}{2})$\\  
\hline
(pair un-fixed, single fixed at 0 or 1) & $\{0,2\}$& $(\frac{1}{2},\frac{1}{2})$\\ 
\hline
(pair un-fixed, single un-fixed) & $\{0,1,2,3\}$& $(\frac{1}{4},\frac{1}{4},\frac{1}{4},\frac{1}{4} )$\\ 
\hline
\end{tabular}

Again, regardless of the partial assignment at the end of Stage 2, the probability of $p$'s \textit{forbidden margin} being realized in Stage 3 is less than or equal to $\frac{1}{2}$. Thus, the probability of $p$ being satisfied (by reaching an \textit{acceptable total}) is greater than or equal to $\frac{1}{2}$.
\end{enumerate}

Thus, for an arbitrary predicate $p\in \mathcal{P}_D$, the probability it is satisfied at the end of Stage 3 is $\geq 1/2$. Since the expectation of the sum is the sum of the expectations, we have that the expected number of predicates satisfied by the random assignment procedure is at least $|\mathcal{P}_D|/2$. As in \citep{RS}, this procedure may be efficiently derandomized using the method of conditional expectations.

Finally, we have a full assignment for the literals $\mathcal{L}$ that satisfies every predicate in $\mathcal{P}_R$ and at least $1/2$ of the predicates in $\mathcal{P}_D$:
\begin{align*}
\text{Total Predicates We Satisfy}\geq|\mathcal{P}_R|+\frac{|\mathcal{P}_D|}{2} &= 
|\mathcal{P}_R|+\frac{(|\mathcal{P}|-|\mathcal{P}_R|)}{2}\\
&= 
\frac{|\mathcal{P}_R|}{2}+\frac{|\mathcal{P}|}{2}\\
&\geq 
\frac{1}{2}\Big(\frac{1}{4}+\frac{5}{2}\epsilon\Big)|\mathcal{P}|+\frac{|\mathcal{P}|}{2}\\
&\geq 
\Big(\frac{5}{8}+\frac{5}{4}\epsilon\Big)|\mathcal{P}|\\
&\geq 
\Big(\frac{5}{8}+\delta\Big)|\mathcal{P}| \hspace{5mm} \text{ for some  $\delta>0$}
\end{align*}

The transition from the second to the third line applies inequality (\ref{quarter}). Since $\epsilon>0$, letting $\delta=\frac{5}{4}\epsilon$ gives the statement about $\delta$ in the final line. 

This reduction runs in polynomial time, and the
final inequality shows that our method exceeds the hardness-of-approximation bound of H{\aa}stad for Max Not-2 (listed as Theorem \ref{hastad} earlier). Thus we have a contradiction. $\Box$.

\subsection{Extended Hardness Result: Gr\"{o}bner Basis Computation for Maximum Degree 2 is Robustly Hard} \label{deg2sec}

In this section we prove the first robust hardness result for Gr\"{o}bner Basis computation for polynomial systems of degree at most 2 (matching the degree bound in the standard NP-hardness result for the tradtional Gr\"{o}bner Basis Problem). Our result is for the $q$-Fractional Gr\"{o}bner Basis Problem defined by Rolnick and Spencer.  

\begin{theorem} \textbf{Robust Hardness For Maximum Degree 2.} \label{extfracharddeg2}
Assume that we are working over a polynomial ring $\mathbb{K}[x_1,x_2,x_3...x_n]$. For any $\epsilon>0$: there is no polynomial-time algorithm $\mathcal{A}$ that solves the $(4/5+\epsilon)$-Fractional Gr\"{o}bner Basis Problem with respect to any lexicographic order (unless P = NP). This statement holds even when $\mathcal{F}$ has maximum degree 2, and each polynomial from $\mathcal{F}$ contains at most 3 variables.
\end{theorem}

Is robust hardness for maximum-degree-2 polynomial systems unique to the $q$-Fractional Gr\"{o}bner Basis Model? For additional perspective, consider that existing hardness results for the $c$-partial Gr\"{o}bner Basis problem of De Leora et al. are based on graph coloring hardness, and are valid for degree bounds that match the chromatic number in the corresponding coloring-hardness result. Since 2-colorings are easy to compute (when they exist), it seems that a non-trivial result for the $c$-partial Gr\"{o}bner Problem in systems of maximum degree 2 would need to take a significantly different approach. 

Our proof of Theorem \ref{extfracharddeg2} is closely inspired by our proof of Theorem \ref{extfrachard}, and will directly reuse much of the notation and language introduced there.  We rely on a (much earlier) hardness result due to H{\aa}stad for a problem involving logical predicates of arity 3 where the system of predicates is satisfiable.  The predicates are now of the following form:
\vspace{-5mm}
\begin{align}\label{oxrform}
OXR(q_1,q_2,q_3)=q_1 \vee (q_2 \oplus q_3)
\end{align}
Here $q_1,q_2,q_3$ are signed literals which represent positive or negated forms of literals from a set $\mathcal{L}$. This clause is true if at least one of $q_1$ or $(q_2 \oplus q_3)$ is true.  The second option $(q_2 \oplus q_3)$ is often called an ``xor" or ``exclusive or." This exclusive or is true when exactly one of $q_2$ or $q_3$ is true. Describing (\ref{oxrform}) above, we will say that $q_1$ is in the \textit{special position} of $p$ and that $q_2$ and $q_3$ are in the \textit{symmetric positions} of $p$. 

\begin{theorem} \label{hastad2}(H{\aa}stad, '01)
For any $\delta>0$, given a satisfiable instance of Max OXR of arity 3, there is no polynomial-time algorithm to find a truth assignment that satisfies a $(\frac{6}{8}+\delta)$-fraction of the predicates (unless $P=NP$). 
\end{theorem}

The fraction above is intentionally not given in lowest form: the unreduced fraction calls attention to the fact that OXR predicates are satisfied by 6 out of 8 of the possible truth settings for the contained literals.

At a structural level, the following proof is very similar to that we gave above for Theorem \ref{extfrachard}, and elimination theory is invoked in the same way.  The differences arise from the form of the logical predicates considered: the pre-processing in Stage 1 is slightly different, the polynomial system constructed has predicate polynomials of lower degree, and the form of these polynomials impacts the analysis of the Gr\"{o}bner-Basis-based partial truth assignment, and the performance of the subsequent coin-flip-based part of the truth assignment.\\

\noindent \textit{Proof (Theorem \ref{extfracharddeg2} ):}  Suppose, for the sake of contradiction, that the $\mathcal{A}$ asserted in Theorem \ref{extfracharddeg2} does exist with $q=(4/5+\epsilon)$ for some fixed $\epsilon>0$. Given an arbitrary satisfiable input $(\mathcal{P}, \mathcal{L})$ of the \textit{Max-OXR Problem of arity 3} we compute an assignment of forbidden quality in polynomial time as follows. Our assignment will be determined over the course of three stages. \\

\noindent \textbf{Stage 1. Instance Tailoring.} Stage 1 removes some predicates and literals from $(\mathcal{P}, \mathcal{L})$ so that certain useful properties hold. 

Iterate through the predicates in $\mathcal{P}$ one at a time.  Consider the signed literals in the symmetric positions of $p$: if both of these signed literals correspond to the same literal, then update $(\mathcal{P}, \mathcal{L})$ according to which of the following cases applies.

\begin{enumerate}
\item If the signed literals in the symmetric positions of $p$ are identical, then their xor must be false (either both of the symmetric-position signed literals are true, or both of the symmetric-position signed literals are false). Thus, in every satisfying assignment the special-position signed literal of $p$ must be true. 
Let $l_i$ denote the literal corresponding to the signed literal in the special position of $p$. Substitute the forced value of $l_i$ into every predicate containing a signed form of $l_i$. Say $l_i$ has been permanently fixed. Remove $l_i$ from $\mathcal{L}$, and remove $p$ from $\mathcal{P}$. 

\item Otherwise the signed literals in the symmetric positions of $p$ are in opposing forms (one positive, one negated). In this case their xor is true for every possible assignment. Remove $p$ from $\mathcal{P}$.
\end{enumerate}
    
   Call the set of all literals fixed during this procedure $\mathcal{L}^f$, and the set of all predicates removed from the original $\mathcal{P}$ by $\mathcal{P}^r$.
After executing the above procedure for every $p\in \mathcal{P}$, observe that $(\mathcal{P}, \mathcal{L})$ now has the following property.  \\

\noindent \textbf{Property 1.} \textit{If $p\in \mathcal{P}$, then the two signed literals in the symmetric positions of $p$ correspond to unique literals.}\\

Literals were only permanently fixed (and removed) when we could reason unequivocally about the truth value they must take in every satisfying assignment. Thus, since the original $(\mathcal{P}, \mathcal{L})$ was satisfiable, the updated $(\mathcal{P}, \mathcal{L})$ is still satisfiable. A predicate was only removed from $\mathcal{P}$ when we could be certain that it would be satisfied by any assignment that extends the partial assignment already constructed for $\mathcal{L}^f$.
    
As in the previous proof, we make a final update to  $(\mathcal{P}, \mathcal{L})$ to remove literals that appear in only one predicate (and the predicates that contain such literals). Call any literal $\l\in \mathcal{L}$ which appears in at most one predicate from $\mathcal{P}$ a \textit{Loner Literal}. Call the set of predicates from $\mathcal{P}$ which contain a Loner Literal by $\mathcal{P}^l$.  We consider temporarily ignoring predicates in $\mathcal{P}^l$ until all non-loner literals have been fixed. For $p\in \mathcal{P}^l$, suppose that the truth values for all literals in $p$ except one loner-literal $l_i$ have already been fixed.\footnote{If more than one signed loner literal in $p$ remains unfixed, then fix all but one arbitrarily, then proceed.} Then:
\begin{itemize}
\item If $l_i$ corresponds to the signed literal in the special position of $p$, then clearly $l_i$ may be set so $p$'s special-position signed literal is true (and $p$ is satisfied). By definition, this choice for $l_i$ (a loner literal) affects no other predicates.
\item If $l_i$ corresponds to a signed literal in a symmetric position of $p$, then, regardless of the truth value of the signed literal in $p$'s other symmetric position\footnote{From Property 1, this other signed literal is not a form of $l_i$.}, there is a truth assignment for $l_i$ that makes $p$'s xor true (so that $p$ is satisfied). By definition, this choice for $l_i$ affects no other predicates. 
\end{itemize}

Since we can successfully satisfy them at the end, we ignore predicates containing loner literals for now: if a predicate contains a \textit{Loner Literal}, remove that predicate from $\mathcal{P}$. Next remove all \textit{Loner Literals} from $\mathcal{L}$. 

Removing predicates may cause additional literals to become \textit{Loner Literals}. Successively remove additional rounds of \textit{Loner}-containing predicates and \textit{Loner Literals}. Mark each \textit{Loner Literal} by the round in which it was removed: once we have created a partial assignment for the remaining system we will fix the values of the \textit{Loner literals} in an order that reverses the order in which they were removed from $\mathcal{L}$ such that all loner-containing predicates are satisfied. 

We now have the $(\mathcal{P}, \mathcal{L})$ that we will argue about for the remainder of the reduction. As in the previous proof, observe that: \\

\noindent \textbf{Property 2.} \textit{After the updates in Stage 1, Every literal $l\in \mathcal{L}$ appears in some form (negated or positive) in at least two predicates from $\mathcal{P}$.}\\

In the proof of Lemma \ref{twofifths}, there was no dependence on the form of the predicates (aside from arity), so again, Property 2 implies that $|\mathcal{P}|\geq\frac{2}{5}(|\mathcal{P}|+|\mathcal{L}|)$.

As in the previous proof, since $100\%$ of the predicates removed from $\mathcal{P}$ in Stage 1 can be polynomial-time satisfied after any partial assignment is fixed for the remaining instance $(\mathcal{P}, \mathcal{L})$, it is sufficient to show that for this remaining instance we can satisfy a $(\frac{6}{8}+\epsilon)$ fraction of the predicates in polynomial time. We construct such an assignment for the remaining literals over two stages.\\

\noindent \textbf{Stage 2. Polynomial Encoding. } Define a polynomial system based on $(\mathcal{P}, \mathcal{L})$ as follows. Create a variable $y_i$ corresponding to the $i$th literal of $\mathcal{L}$. Denote this set of variables by $Y$. Create a polynomial $y_i(1-y_i)$. This gives a set of $|\mathcal{L}|$ ``literal polynomials" whose mutual roots are exactly $\{0,1\}^{|\mathcal{L}|}$. 

Next create a set of predicate polynomials. Consider $p\in\mathcal{P}$.
We create a polynomial corresponding to $p$ as follows: it will be the product of 2 terms.  The first term will correspond to $p$'s special position. If the signed literal in the special position of $p$ corresponds to literal $l_i$ and appears in positive form, then the first term in $p$'s polynomial will be $(y_i-1)$. If the signed literal in the special position of $p$ corresponds to literal $l_i$ and appears in negated form, then the first term in $p$'s polynomial will be $(y_i)$. The second term in $p$'s polynomial will correspond to $p$'s xor.  For $p$s xor to be true, there is a single \textit{acceptable sum} for variables corresponding to $p$s symmetric-position signed literals. Let $l_j$ and $l_k$ denote the literals corresponding to the signed literals in the symmetric positions of $p$.  We summarize the construction of the second term of $p$s polynomial below:\\
\begin{tabular}{l|l}
Form of $p$s xor & Form of second term in product-defined polynomial for $p$\\ \hline
$(l_j\oplus l_k)$ & $(y_j+y_k-1)$ \\
$(\neg l_j\oplus l_k)$ & $((1-y_j)+y_k-1)$ \\
$(l_j\oplus \neg l_k)$ & $(y_j+(1-y_k)-1)$ \\
$(\neg l_j\oplus \neg l_k)$ & $((1-y_j)+(1-y_k)-1)$\\
\end{tabular}

\vspace{4mm}
The product of the two described terms gives the polynomial for $p$. Note that $l_j$ and $l_k$ must be different literals from Property 2, but that $i$ might match one of $j$ or $k$.  Since the literals in $\mathcal{L}^f$ already have permanently-fixed truth values, the corresponding 0s or 1s are substituted into the predicate polynomials defined above.  
For example, the predicate $\neg l_i \vee (l_j\oplus \neg l_k)$ gives polynomial 
\[
y_i(y_j+(1-y_k)-1)= y_i(y_j-y_k).
\]
For further example, if $l_i$ was set to 1 in Stage 1, and $l_j$ and $l_k$ remain unfixed, then $\neg l_i \vee (l_j\oplus \neg l_k)$ would produce the simple predicate polynomial
\[
y_j-y_k.
\]

This constructs a set of predicate polynomials of cardinality $|\mathcal{P}|$.
Each predicate polynomial is the product of two terms each of degree at most 1: each predicate polynomial has maximum total degree at most 2. Also, as in the previous reduction, each predicate polynomial contains at most 3 variables (corresponding to a limit of three signed literals per OXR

Let $\mathcal{F}$ denote the system of polynomials containing both the literal polynomials and the predicate polynomials. By our construction, every satisfying assignment for $(\mathcal{P},\mathcal{L})$ can be interpreted as a point in the variety defined by $\mathcal{F}$. In particular, since $(\mathcal{P},\mathcal{L})$ is satisfiable, $V(\langle \mathcal F \rangle)$ is non-empty.

Apply algorithm $\mathcal{A}$ to solve the $q$-Fractional Gr\"{o}bner Basis Problem for $\mathcal{F}$ for $q=(4/5+\epsilon)$ for some fixed $\epsilon>0$. Let $Y'$ denote the variables that $\mathcal{A}$ selects to ignore. The set $Y'$ was chosen so that
\begin{align*}
|\mathcal{F}_{Y'}|\leq (1-q)|\mathcal{F}|&\leq (1/5-\epsilon)|\mathcal{F}|\\
&\leq (1/5-\epsilon)(|\mathcal{P}|+|\mathcal{L}|)\\
&\leq (1/5-\epsilon)\Big(\frac{5}{2}|\mathcal{P}|\Big)\\
&\leq \Big(\frac{1}{2}-\frac{5}{2}\epsilon\Big)|\mathcal{P}|
\end{align*}

The third line follows from the second line due to Lemma \ref{twofifths}. 

Let $\mathcal{P}_D$ denote the set of predicate polynomials that are in $\mathcal{F}_{Y'}$, and $\mathcal{P}_R$ denote the set of predicate polynomials that are in $\mathcal{F}\backslash\mathcal{F}_{Y'}$. Clearly $|\mathcal{P}_D|\leq |\mathcal{F}_{Y'}|\leq \Big(\frac{1}{2}-\frac{5}{2}\epsilon\Big)|\mathcal{P}|$, so:
\begin{align}\label{half}
|\mathcal{P}_R|= |\mathcal{P}|- |\mathcal{P}_D| \geq |\mathcal{P}|- \Big(\frac{1}{2}-\frac{5}{2}\epsilon\Big)|\mathcal{P}|= \Big(\frac{1}{2}+\frac{5}{2}\epsilon\Big)|\mathcal{P}|
\end{align}

That is, $\mathcal{A}$ computes a Gr\"{o}bner Basis with respect to a lexicographic order for the ideal generated by the polynomials in $\mathcal{F}\backslash \mathcal{F}_{Y'}$, and inequality (\ref{half}) says that at least a $(\frac{1}{2}+\frac{5}{2}\epsilon)$ fraction of the predicate polynomials must be in $\mathcal{F}\backslash \mathcal{F}_{Y'}$. The satisfiability of $(\mathcal{P}, \mathcal{L})$ ensured that $V(\langle \mathcal{F}\rangle)$ was non-empty, so $V(\langle \mathcal{F}\backslash \mathcal{F}_{Y'}\rangle)$ is certainly non-empty.

Given the Gr\"{o}bner Basis for $\langle\mathcal{F}\backslash \mathcal{F}_{Y'}\rangle$ with respect to a lexicographic order, a point in the variety of $\langle\mathcal{F}\backslash \mathcal{F}_{Y'}\rangle$ can be efficiently computed via successive elimination of the variables: since the variety is finite (it is a subset of $\{0,1\}^{|Y\backslash Y'|})$, all partial solutions extend, and for each successive variable elimination only 2 options must be checked to find some $y_i$ that works. This point in the variety is a vector $y$ of length $|Y\backslash Y'|$ which is a mutual zero of all polynomials in $\mathcal{F}\backslash \mathcal{F}_{Y'}$. Each entry in the vector corresponds to some literal variable in our Max OXR instance:
if $y_i$ is 1 in this vector, assign the corresponding literal $l_i$ to be True, if $y_i$ is 0 in this vector, assign the corresponding literal $l_i$ to be False. Since $y\in \{0,1\}^{|Y\backslash Y'|}$, this routine makes an assignment for every $l_i$ corresponding to a  $y_i \in Y\backslash Y'$. 

The vector $y$ is a mutual zero of polynomials in $\mathcal{F}\backslash \mathcal{F}_{Y'}$: substituting $y$ into any predicate polynomial
in $\mathcal{F}\backslash \mathcal{F}_{Y'}$ gives 0. Consider the factored form of the predicate polynomials we constructed: such a polynomial evaluates to 0 under our constructed truth assignment if, and only if, either:
\begin{itemize}
\item 
the signed literal corresponding to $p$s special position is True or 
\item the xor over $p$s symmetric-position signed literals is True (or both). 
\end{itemize}
Thus, since substituting $y$ into any predicate polynomial
in $\mathcal{F}\backslash \mathcal{F}_{Y'}$ gives 0, our $y$-based partial truth assignment satisfies every predicate whose polynomial is in $\mathcal{P}_R$.\\

\noindent \textbf{Stage 3. Supplemental Random Assignment.} The literals corresponding to variables in $Y'$ have not yet been assigned truth values. For these literals, consider a random independent-fair-coin-flip procedure that assigns the literal to be True with probability $1/2$, and False with probability $1/2$. We argue that regardless of the partial truth assignment constructed in Stage 2 (and effectively in Stage 1), in expectation this random procedure satisfies at least half of the predicates corresponding to the polynomials in $\mathcal{P}_D$. Such a procedure can be derandomized via the method of conditional expectations to obtain a deterministic assignment algorithm with quality that matches the expected value.

Let $p$ denote a OXR predicate corresponding to a polynomial in $\mathcal{P}_D\subseteq \mathcal{F}_{Y'}$. From the form of $\mathcal{F}_{Y'}$,  $p$ contains (some form of) at least one literal corresponding to a $y_i\in Y'$: before the coin-flip procedure, the truth value of such a literal has not yet been decided. Through case analysis, we show that the probability that $p$ is satisfied at the end of the coin-flip procedure is always at least $1/2$.
Before the coin-flip procedure, suppose that:
\begin{itemize}
\item The truth value of the signed literal in $p$s special position has not yet been decided. The coin-flip procedure sets this special-position signed literal to be true with probability $1/2$. Thus, the probability that $p$ is true at the end of the coin-flip procedure is at least $1/2$. 
\item Otherwise, the truth value of the signed literal in $p$s special position has already been decided. Then either:
\begin{itemize}
\item Exactly one of the signed literals in a symmetric position of $p$ has not yet been decided. Since the signed literal in $p$'s other symmetric position has a fixed value, the probability that $p$s xor over the two symmetric positions is true as a result of the coin-flip procedure is $1/2$. Thus, the probability that $p$ is true at the end of the coin-flip procedure is at least $1/2$. 
\item Otherwise, both of the signed literals in the symmetric positions of $p$ have not yet been decided. From Property 2, these symmetric-position signed literals correspond to unique literals. Thus, the probability that $p$s xor over the two symmetric positions is true as a result of the coin-flip procedure is $1/2$ ($p$s xor is satisfied by exactly half of the possible truth assignments). Thus, the probability that $p$ is true at the end of the coin-flip procedure is at least $1/2$. 
\end{itemize}
\end{itemize}
Thus, for an arbitrary predicate $p\in \mathcal{P}_D$, the probability it is satisfied at the end of Stage 3 is $\geq 1/2$. Since the expectation of the sum is the sum of the expectations, we have that the expected number of predicates from  $\mathcal{P}_D$ that are satisfied by the random assignment procedure is at least $|\mathcal{P}_D|/2$. This procedure may be efficiently derandomized using the method of conditional expectations.

Finally, we have a full assignment for the literals $\mathcal{L}$ that satisfies every predicate in $\mathcal{P}_R$ and at least $1/2$ of the predicates in $\mathcal{P}_D$:
\begin{align*}
\text{Total Predicates We Satisfy}\geq|\mathcal{P}_R|+\frac{|\mathcal{P}_D|}{2} &= 
|\mathcal{P}_R|+\frac{(|\mathcal{P}|-|\mathcal{P}_R|)}{2}\\
&= 
\frac{|\mathcal{P}_R|}{2}+\frac{|\mathcal{P}|}{2}\\
&\geq 
\frac{1}{2}\Big(\frac{1}{2}+\frac{5}{2}\epsilon\Big)|\mathcal{P}|+\frac{|\mathcal{P}|}{2}\\
&\geq 
\Big(\frac{6}{8}+\frac{5}{4}\epsilon\Big)|\mathcal{P}|\\
&\geq 
\Big(\frac{6}{8}+\delta\Big)|\mathcal{P}| \hspace{5mm} \text{ for some  $\delta>0$}
\end{align*}

The transition from the second to the third line applies inequality (\ref{half}). Since $\epsilon>0$, letting $\delta=\frac{5}{4}\epsilon$ gives the statement about $\delta$ in the final line. 

This reduction runs in polynomial time, and the
final inequality shows that our method exceeds the hardness-of-approximation bound of H{\aa}stad for Max OXR (listed as Theorem \ref{hastad2} earlier). Thus we have a contradiction. $\Box$.


\section{Conditional Results for the Strong $c$-partial Gr\"{o}bner Basis Problem}

In addition to problems which have been proved to be NP-Hard, there are problems which are conjectured to be NP-hard. Some of these conjectures have special significance because they can be used to prove conditional hardness-of-approximation results that perfectly match the best existing positive algorithmic results for famous combinatorial optimization problems. Khot's seminal work on the Unique Games Conjecture (\cite{Khot02}) included several conjectures on the NP-Hardness of specialized variants of the Label Cover Problem (for this body of work, Khot was awarded the 2014 Rolf Nevanlinna Prize by the International Mathematical Union). These conjectures have been closely scrutinized but little evidence has arisen against them \cite{dinur}. In this section we give several results that depend on conditional results of Dinur, et al \cite{dinur}.

Given polynomial system $\mathcal{F}$ on variable set $X$, call $Y\subseteq X$  an \textit{independent set of variables} if no pair of variables in $Y$ appear in a common polynomial from $\mathcal{F}$.
Recall the Strong $c$-partial Gr\"{o}bner Problem as defined by De Loera, et al.:

\begin{definition} \citep{deloera}
Define the \textbf{Strong $c$-partial Gr\"{o}bner Problem} as follows. Given as input, a set $\mathcal{F}$ of polynomials on a set $X$ of variables, output the following:
\begin{itemize}
\item disjoint $X_1,...,X_b\subseteq X$, such that $b\leq c$ and each $X_i$ is an independent set of variables,
\item $X'\subseteq X$ where $X'=X\backslash (\cup_{i=1}^b X_i)$ (i.e., we have taken away at most $c$ independent sets of variables),
\item $\mathcal{F}'\subseteq\mathcal{F}$ such that $\mathcal{F}'$ consists of all polynomials in $\mathcal{F}$ involving only variables in $X'$,
\item a Gr\"{o}bner basis for $\langle \mathcal{F}'\rangle$ over $X'$ (where the monomial order of $X$ is restricted to a monomial order on $X'$).
\end{itemize}
\end{definition}

Existing hardness guarantees for this problem are parametrized statements: in both \citep{deloera} and \citep{RS} the $c$ for which NP-hardness is proved grows only linearly with the degree bound for $\mathcal{F}$. De Loera et al. give such linear relationship with slope of roughly 2/3 that is valid starting from surprisingly low degree bounds (polynomial systems of degree at most 3). In contrast, Rolnick and Spencer show that a linear relationship of arbitrarily high slope becomes valid above some sufficiently-high degree bound. Is linear dependence of $c$ on the degree bound for $\mathcal{F}$ necessary for NP-Hardness to hold? The answer may be ``no" in quite a dramatic sense. We offer two conditional results:

\begin{theorem} \label{Khotcond}(Hardness conditioned on Khot)
If Khot's $2\leftrightarrow 2$ conjecture is true, then for arbitrary constant $k$, the following problem is NP-Hard:
solve the Strong $k$-partial Gr\"{o}bner Problem for some lexicographic order.
This statement holds even if $\mathcal{F}$ has maximum degree 4.
\end{theorem}

\begin{theorem} \label{Dinurcond}(Hardness conditioned on Dinur et al.)
If Dinur et al.s $\mytie$-shaped conjecture is true, then for arbitrary constant $k$, the following problem is NP-Hard:
solve the Strong $k$-partial Gr\"{o}bner Problem for some lexicographic order.
This statement holds even if $\mathcal{F}$ has maximum degree 3.
\end{theorem}

To prove these conditional results, we use the reduction technique from \citep{RS} for the large-linear-slope case of the Strong $c$-partial Gr\"{o}bner Basis Problem as a template. That proof uses a '94 NP-hardness result of Lund and Yanakakis \citep{Lundyan} for a certain form of decision problem in coloring known as the Approximate Coloring Problem. 
\textsc{ApproxColoring}$(q,Q)$ is the following decision problem: given a graph $G$ decide between the following two alternatives:
\begin{itemize}
\item $\chi (G)\leq q$ , or
\item $\chi (G)\geq Q$.
\end{itemize}

Lund and Yanakakis showed that for any $h>1$, there exists a $k$ for which \textsc{ApproxColoring}$(k,hk)$ is NP-Hard.  Rolnick and Spencer use the ability to solve the Strong $c$-partial Gr\"{o}bner Basis Problem closely as a black box to efficiently decide between these two alternatives. If $G$ is $k$-colorable, their polynomial-time method outputs a proper coloring of $G$ by $hk-1$ colors. Such a coloring is clearly impossible if $\chi (G)\geq hk$, so the output makes clear which alternative holds.

In 2009, Dinur, Mossel and Regev proved in \citep{dinur} that conjectures about the hardenss of certain Label Cover problems have dramatic implications for the \textsc{ApproxColoring}$(q,Q)$ Problem. Substituting the following conditional results due to Dinur et al. in the place of the result of Lund and Yanakakis result in the reduction from \citep{RS} immediately gives Theorems $\ref{Khotcond}$ and $\ref{Dinurcond}$ respectively.

\begin{theorem} (Dinur et al. '09)
For any constant $Q>4$, the $2\leftrightarrow 2$ conjecture of Khot implies that \textsc{ApproxColoring}$(4,Q)$ is NP-Hard. This also holds for \textsc{ApproxColoring}$(q,Q)$ for any $q\geq 4$.
\end{theorem}

\begin{theorem} (Dinur et al. '09)
For any constant $Q>3$, the $\mytie$ conjecture implies that \textsc{ApproxColoring}$(3,Q)$ is NP-Hard. This also holds for \textsc{ApproxColoring}$(q,Q)$ for any $q\geq 3$.
\end{theorem}

\section{Directions and Conclusion}

We have argued about robust hardness only with respect to lexicographic orders. Empirically, other term orders often result in much faster compute times for Buchberger's Algorithm. For example, during Buchberger's Algorithm, lexicographic orders have been shown to sometimes produce very-large-degree intermediate polynomials, while other orders (e.g. revlex, grevlex) have fewer such problems. For other term orders, are there analogs of the hardness results here and in \citep{deloera}, \citep{RS}?

We conclude with a comment.  The family of hardness results our proofs are based on (and also that \citep{RS} is based on) represent each member of the family of arity-three logical predicate problems which H{\aa}stad identified as the arity-3 cases for which the very simple random-coin-flip-based truth assignment achieves an approximation ratio that is tight (even for the case of satisfiable instances). In particular, a simple random assignment matches the hardness results of H{\aa}stad stated in this paper as Theorems \ref{hastad} and \ref{hastad2}: deciding truth values by fair coin flips gives a randomized $(5/8)$-approximation algorithm for Max Not-2, a randomized $(6/8)$-approximation algorithm for Max OXR, and a randomized $(7/8)$-approximation algorithm for Max 3-SAT. H{\aa}stad proved $(6/8)$-tightness for Max OXR and $(7/8)$-tightness Max 3-SAT in 2001 \citep{hast01}, but only completed the family of results in 2014 by proving $5/8$ is tight for Max Not-2 \citep{hast14}.

By viewing this family of satisfiability problems in terms of the strength of the Robust-Gr\"{o}bner-Basis results that their tight hardness-of-approximation guarantees can be used to produce, we have a new way to understand the relationships between them. Referring back to Figure \ref{graphofq}, for example, we see a rather surprising message: though it took 13 additional years for H{\aa}stad to prove that $5/8$ is truly tight for Max Not-2 on satisfiable instances, our Robust-Gr\"{o}bner-Basis view shows that there is a sense in which his 2001 result for Max OXR on satisfiable instances (with weaker parameter $6/8$) is actually stronger. In particular, from an algebraic perspective, our Max-OXR based result for maximum degree 2 (our Theorem \ref{extfracharddeg2}) may seem like a much more surprising statement on Robust Hardness of Gr\"{o}bner-Basis computation (despite the weaker $q$ parameter) than our stronger-parameter result for maximum degree 3 (Theorem \ref{extfrachard}).

\bibliographystyle{plain}
\bibliography{references}
\end{document}